# Optimal Final Carry Propagate Adder Design for Parallel Multipliers

Ramkumar B and Harish M Kittur

*Abstract*- **Based on the ASIC layout level simulation of 7 types of adder structures each of four different sizes, i.e. a total of 28 adders, we propose expressions for the width of each of the three regions of the final Carry Propagate Adder (CPA) to be used in parallel multipliers. We also propose the types of adders to be used in each region that would lead to the optimal performance of the hybrid final adders in parallel multipliers. This work evaluates the complete performance of the analyzed designs in terms of delay, area, power through custom design and layout in 0.18 um CMOS process technology**.

## I. INTRODUCTION

The critical signal path in a parallel multiplier is divided into three domains: AND gate array, PPST (Partial Product Summation Tree) and the final CPA. The delay introduced by the AND gate is relatively small compared to the other two components, especially for the large size multiplier. This delay component is also relatively independent of the size of the multiplier. The delay introduced by the PPST and the final CPA constitutes a dominant component of the delay in the multiplier [1].

Hybrid CPA have been proposed earlier with detailed investigations on the final addition of parallel multipliers [1]-[3]. It is well known that the signals applied to the inputs of the CPA arrive first at the ends of the CPA and the last ones are those in the middle of the CPA. So the determination of the exact arrival time to final adder is of prime importance in the design of the optimal final adder. We have therefore analyzed the arrival time from the PPST through layout implementation and based on those arrival times, the inputs has been applied to the 7 type of adders for 4 different bit sizes ( total of 28 adders). The analysis is done by using industry standard tool and based on the post layout simulation results we have designed the optimal final structure. The investigation includes 8 by 8, 16 by 16, 32 by 32 and 64 by 64 Dadda multiplier with the final adders being 16, 32, 64 and 128-bit Ripple Carry Adder (RCA), Carry Save Adder (CSA), Carry Select Adder (CSLA), Carry Look Ahead Adder (CLA) and BEC (Binary to Excess1) based adders called here as BEC Carry Select Adder (BCSLA), BEC Carry Save Adder (BCSA) and BEC Carry Look Ahead adder (BCLA)[4]-[10].

This paper is structured as follows; Section II deals with the design of the PPST based on Dadda algorithm and analysis of the signal arrival profile from the PPST. The analysis of the performance of various adders in terms of area, delay and power is in Section III. The equations for efficient partitioning of the multiplier region are developed in Section IV. The final adder design and ASIC implementation details are provided in Section V and VI respectively. Finally the work is concluded in Section VII.

This work was supported in part by the Integrated Circuit Design Laboratories, VIT University, Vellore, India.

B.Ramkumar[1] and Harish Kittur[2] are with the VLSI division, School of Electronics Engineering, VIT University, Vellore, India. (email: ramkumar.b@vit.ac.in[1]; kittur@vit.ac.in[2])

## II. ANALYSIS OF PPST SIGNAL ARRIVAL PROFILE

The basic top-level implementation for $N$ by $N$ unsigned parallel multiplier without CPA is shown in Fig. 1. To analyze the exact arrival time from the PPST, the multiplier is implemented without CPA.

### *Signal Buffering*

In order to determine typical signal arrival profile and drive strengths, D flip-flops are used on the primary inputs & outputs. D flip-flops drive multiple buffers to distribute input signals to $N^2$ AND gates. Delay simulations were performed for each cell library to resolve,

a) The maximum number of buffers that a single D flip-flop can drive.

b) The maximum number of AND gate inputs that a single buffer can drive.

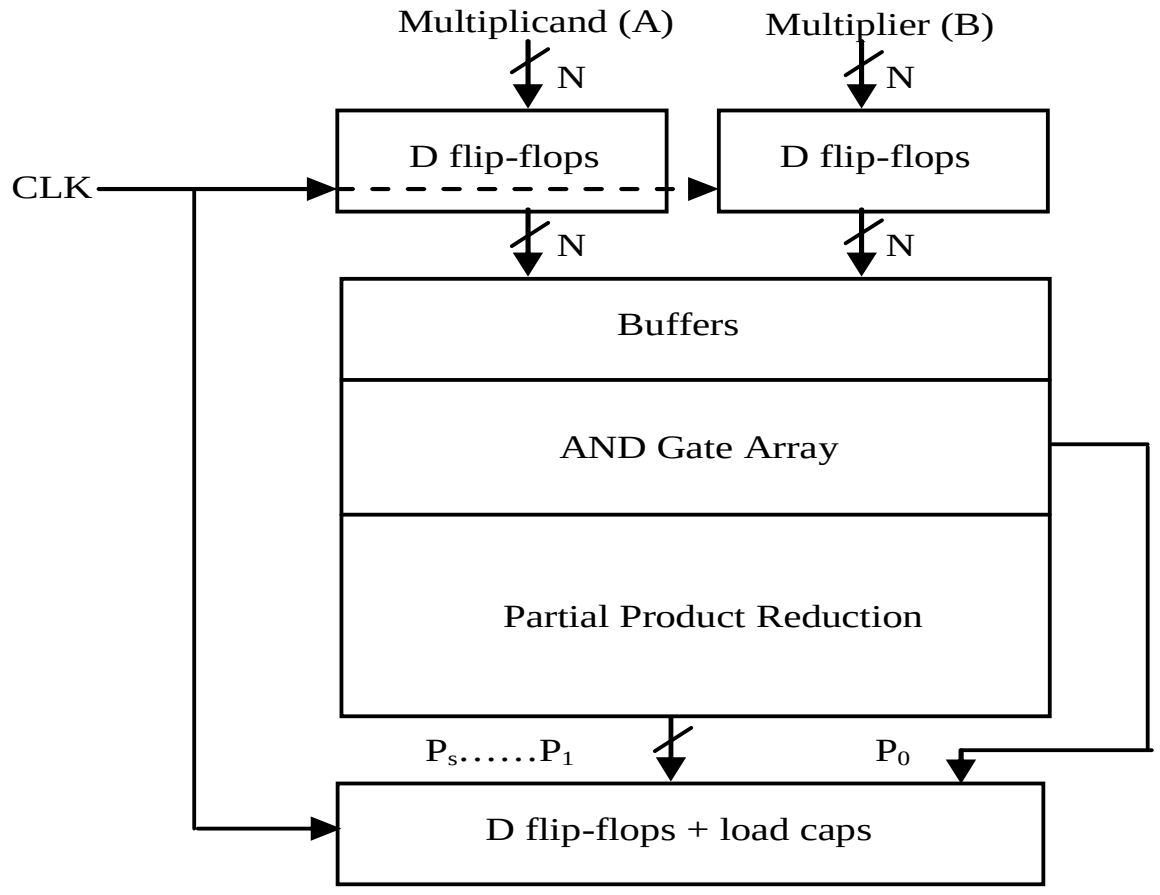

Fig. 1. Top-level implementation of $N$ by $N$ multiplier without CPA

*Partial Product Reduction Tree*

The Wallace and Dadda methods are the popular partial product reduction algorithm for fast multipliers [11]-[12], but a closer examination of the delays and area within these two multiplier shows that the Dadda multiplier is slightly faster and area efficient than the Wallace multiplier [13]-[14].

For the reduction of the $N$ by $N$ partial matrix, Dadda proposes a sequence of matrix heights that are determined by working back from the final two-row matrix. In order to implement the minimum number of reduction stages, the height of each intermediate matrix is limited to the largest integer that is no more than 1.5 times the height of its successor. In this paper, the PPST is implemented based on the Dadda algorithm.

## III. FINAL ADDER ANALYSIS

Once the partial product matrix has been reduced to a height of two, the final stage CPA length is determined in the Dadda approach as below,

$$\text{CPA length} = 2N - 2$$

Fig.1 details the connection of a AND gate array (p0) and compression strategy (p1 to ps) to a D flip-flop and capacitive load which finds the arrival time from the inputs to the final CPA. The evaluated arrival profile is fixed as input delay to the CPA using timing constraint file as shown in Fig. 2.

### *A. 8 by 8 Dadda multiplier*

The input arrival profile to the final CPA of 8 by 8 Dadda multiplier is shown in Fig. 3a. Based on the arrival profile it is divided into 3 regions, in which the first region has a positive slope from the point 0 to 5 and the second region has a constant region from point 6 to 14 and the third region has a negative slope which is point 15. Since the negative slope has only one point we can include this region within the second region.

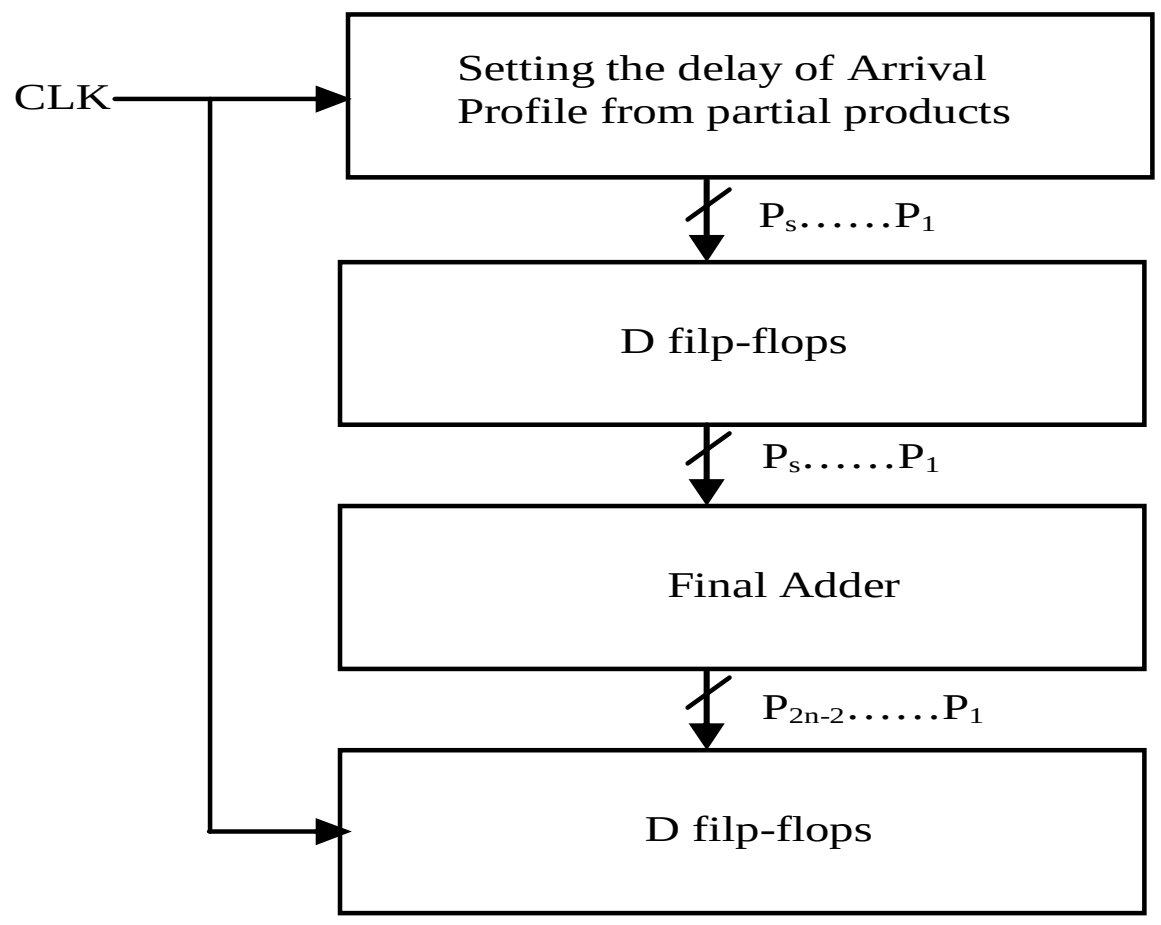

Fig. 2. Arrival profile evaluation to the CPA for $N$ by $N$ multiplier

The performance comparisons, of the 7-types of 16-bit adders for the 8 by 8 multiplier, in terms of output timing are shown in Fig. 3(b) and Fig. 3(c). The area and power comparison is shown in Fig. 3(d) and Fig. 3(e) respectively. These values are predicted from the post layout simulation of each adder. In each of the 3 region, we note that more than one adder is performing well with only slight differences. In the positive slope region of the comparison graph, the RCA and CSA are performing faster. The difference between the arrival times of each successive point in positive slope is greater than the carry propagation time of one full adder of an RCA. i.e., consider a full adder takes 0.24 ns to propagate the carry. But the difference between the arrival times of successive points in positive slope is more than 0.24 ns. So the RCA is sufficient in this region. Also due to the greater area and power requirement of CSA, the RCA is the best choice for the positive slope region.

In the second region the arrival profile of the multiplier signals are relatively constant. So these bits are waiting for the carry inputs from the LSB side. Thus a faster adder is needed in this region. The BCSA and BCLA perform faster in this region, shown in Fig 3b. On comparing BCSA with BCLA, the BCLA is slightly faster than the BCSA and also the area and power of the BCLA are lesser than the BCSA. So we conclude that for the 8 by 8 multiplier the hybrid adder should have RCA for the region 1 and BCLA for the region 2.

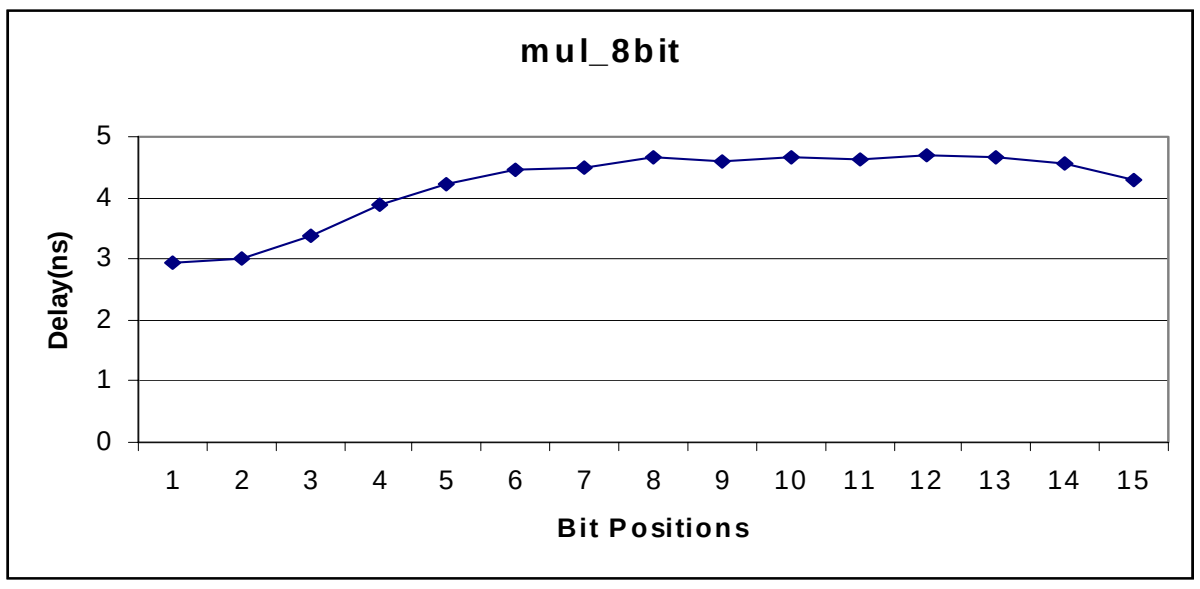

(a)

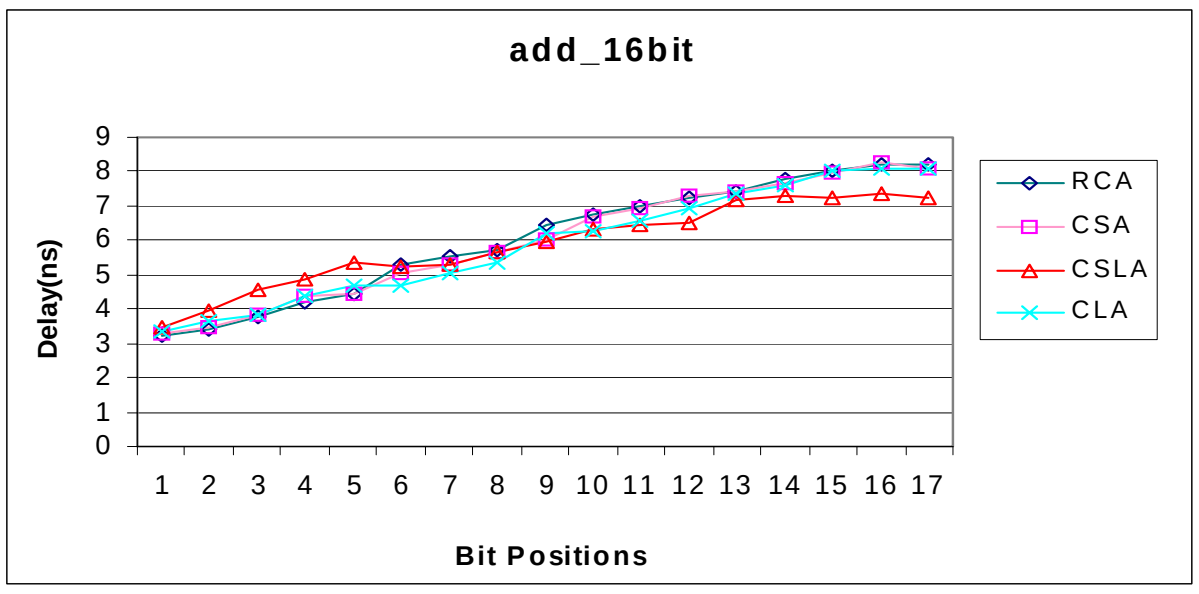

(b)

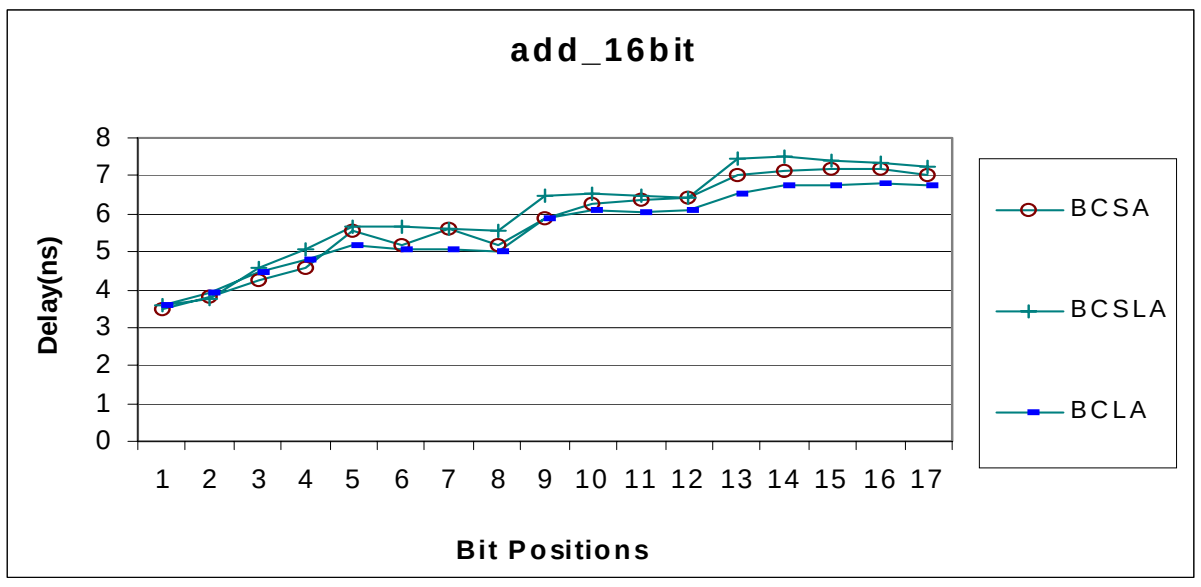

(c)

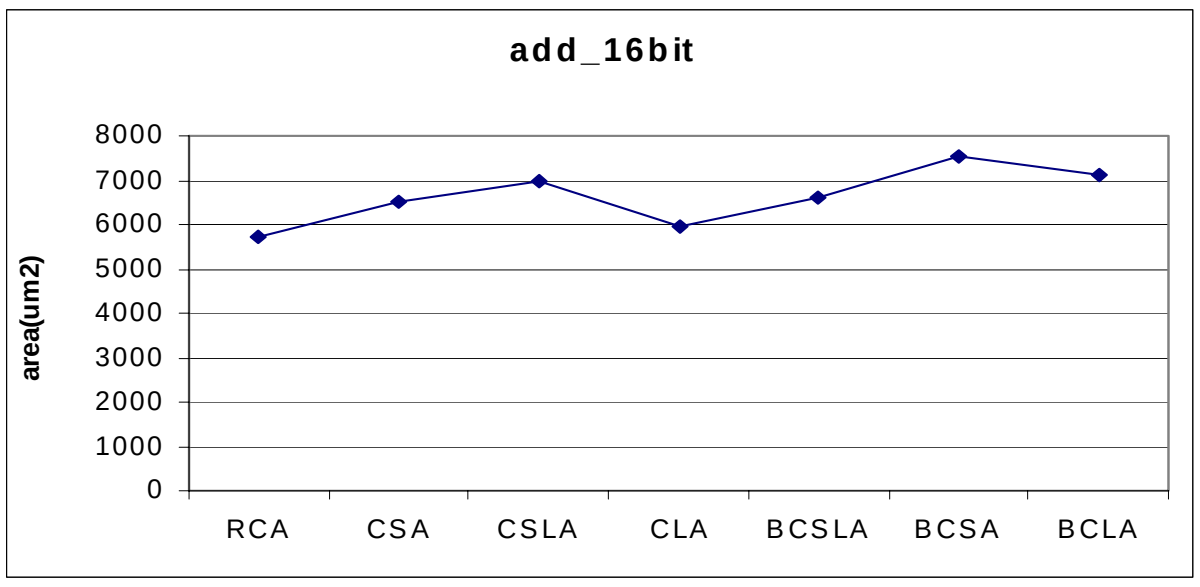

(d)

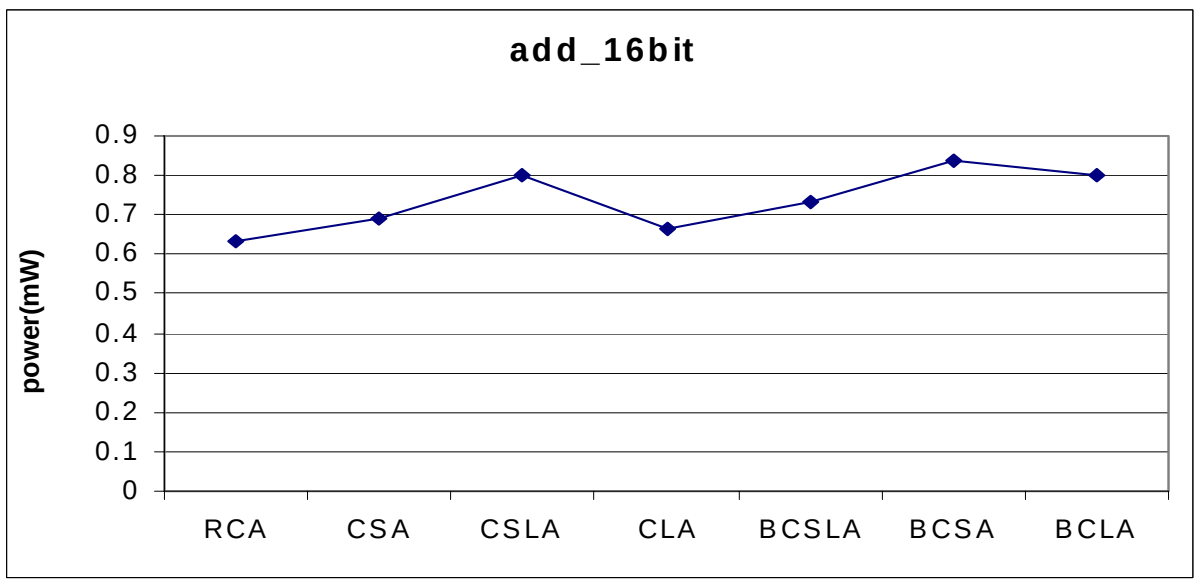

(e)

Fig. 3. Final adder analysis for 8 by 8 Dadda multiplier: (a) arrival profile of PPST, (b) delay analysis of RCA, CSA, CSLA and CLA, (c) delay analysis BCSA, BCSLA and BCLA, (d) area analysis, (e) power analysis.

## *B. 16 by 16 Dadda multiplier*

The input arrival profile for the final adder of 16 by 16 Dadda multiplier is shown in Fig. 4(a). Based on the arrival profile, here also we can divide this in to 3 regions, in which the first region has a positive slope from the point 0 to 6 and the second region has a constant region from point 7 to 25 and the third region has a negative slope from the point 26 to 31. Since the negative slope region has more than a few bits we consider this a separate region.

The performance comparisons, of 7-types of 32-bit adders for the 16 by 16 multiplier, in terms of output timing are shown in Fig. 4(b) and Fig. 4(c). The area and power comparisons are shown in Fig. 4(d) and Fig. 4(e) respectively. In the positive slope region, here also the carry propagation time of one full adder is smaller than the arrival inputs between successive points of the positive slope region. Thus the RCA is again the best choice in this positive slope region. In the second region, we find three adders are working faster with only slight difference. They are CSLA, BCSA and BCSLA. The BCSA works faster than the CSLA and BCSLA in this region. But due to increase in the area and power of the BCSA, we can choose one of the adders from the CSLA and BCSLA for the second region. On comparing these two adders, the CSLA is slightly faster than the BCSLA and it leads to more area and power than the BCSLA. So we suggest that the BCSLA be used for the second region. In the third region, the BCSA and BCLA are working faster. The BCLA saves more area and power than the BCSA; we can therefore choose the BCLA for the entire third region.

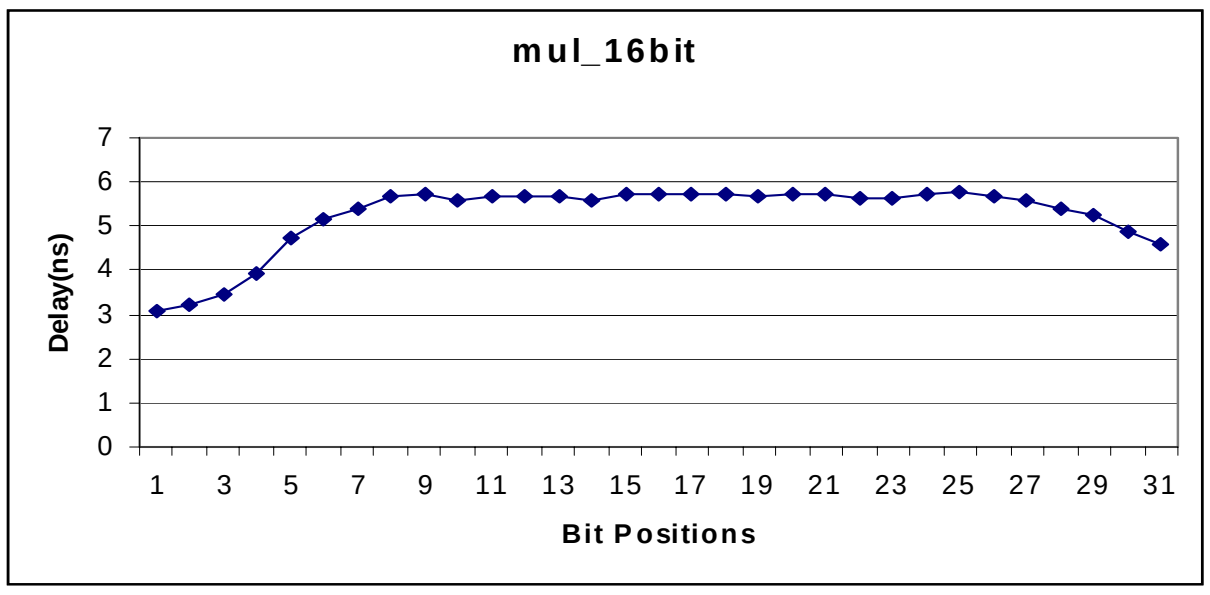

(a)

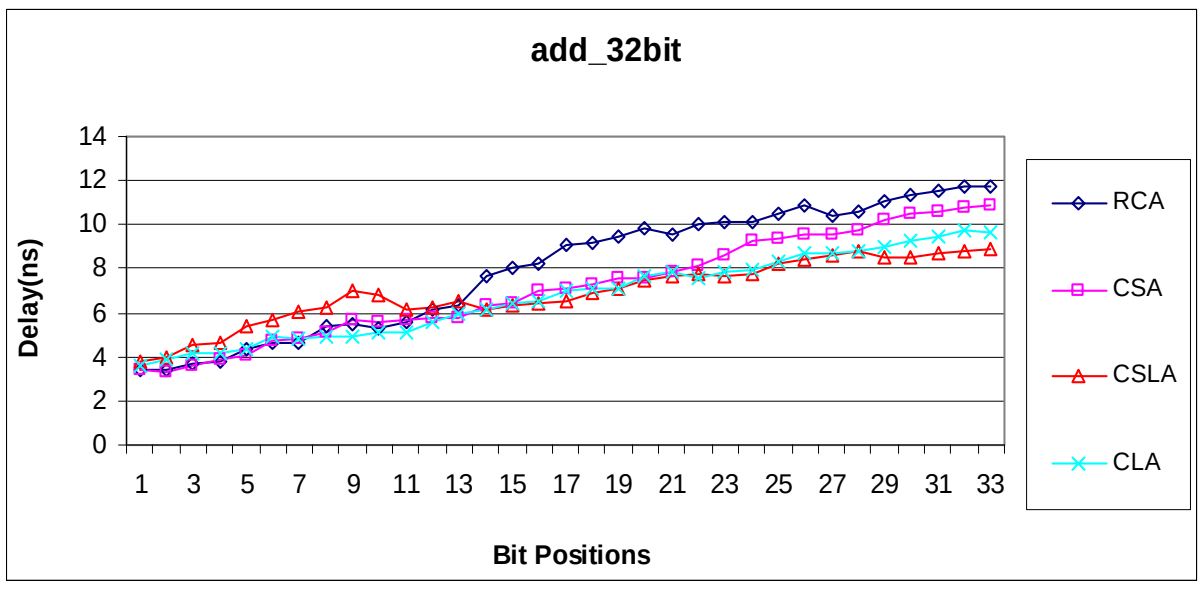

(b)

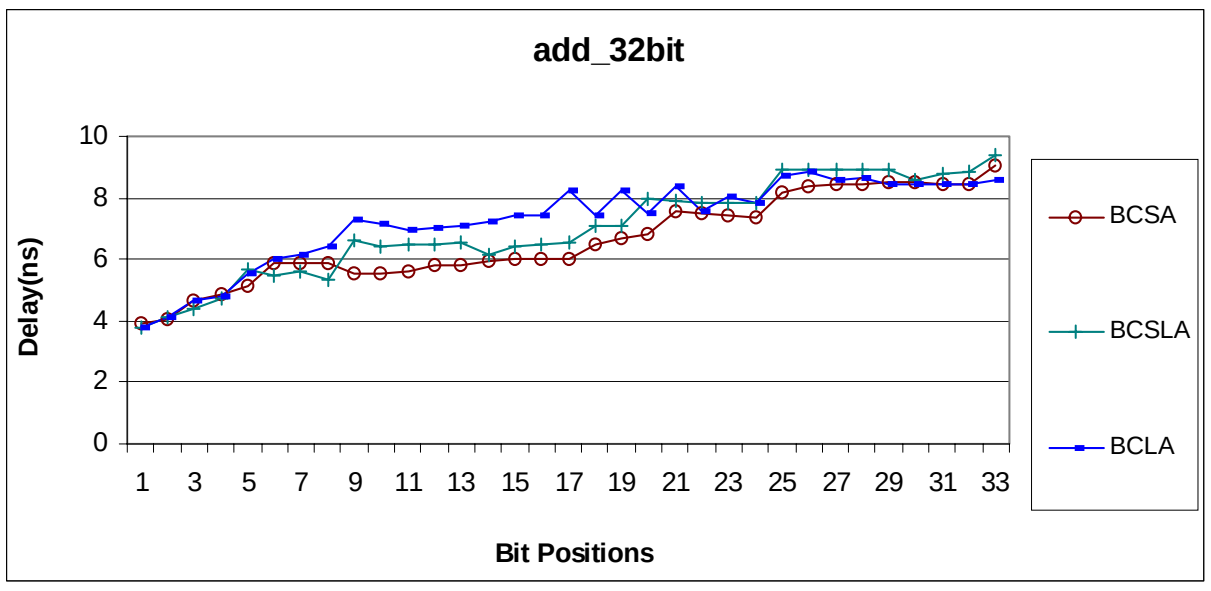

(c)

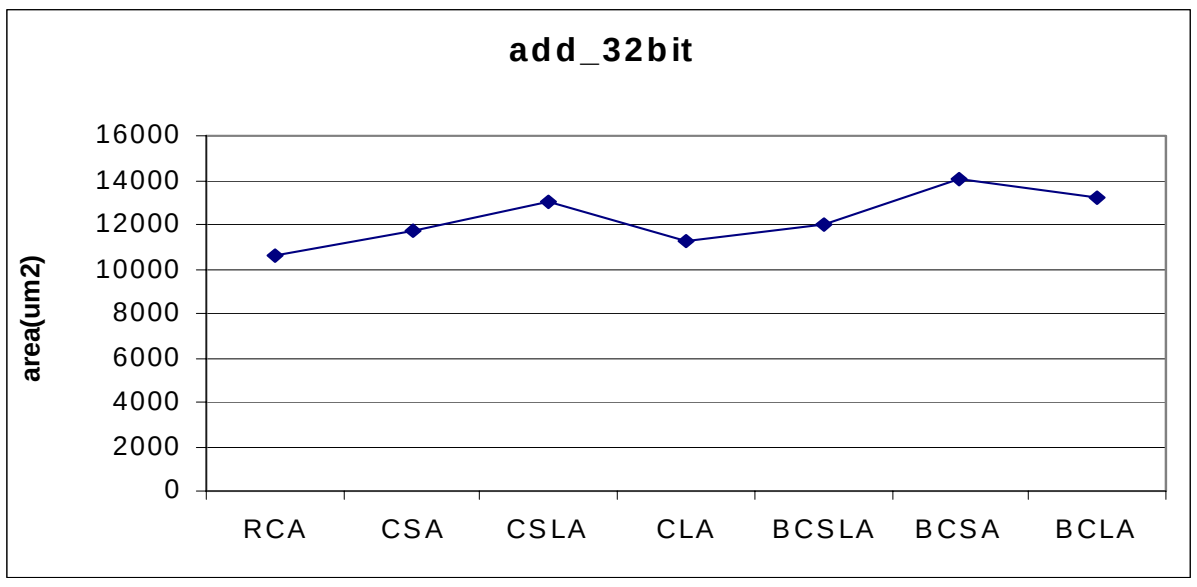

(d)

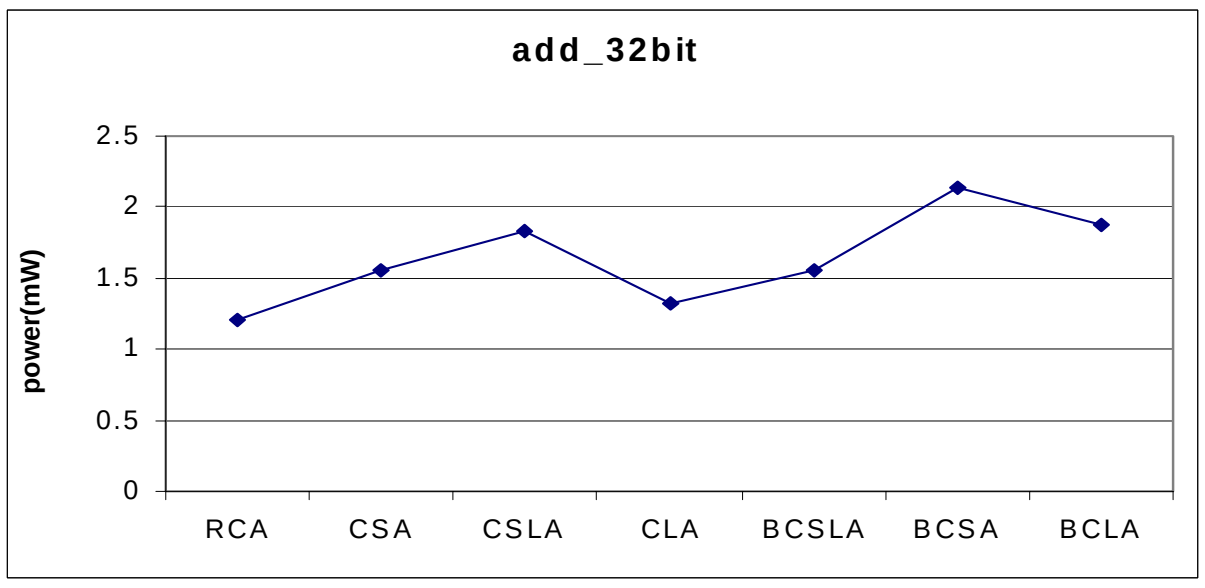

(e)

Fig. 4. Final adder analysis for 16 by 16 Dadda multiplier: (a) arrival profile of PPST, (b) delay analysis of RCA, CSA, CSLA and CLA, (c) delay analysis BCSA, BCSLA and BCLA, (d) area analysis, (e) power analysis.

### *C. 32 by 32 Dadda multiplier*

The input arrival profile to the final adder of 32 by 32 Dadda multiplier is shown in Fig. 5(a). Based on the arrival profile, here also we can divide this in to 3 regions, in which the first region has a positive slope from the point 0 to 15 and the second region has a constant region from point 16 to 52 and the third region has a negative slope from the point 53 to 63.

The performance comparison of 7-types of 64-bit adders for the 32 by 32 multiplier in terms of output timing is shown in Fig. 5(b) and Fig. 5(c). The area and power comparison is shown in Fig. 5(d) and Fig. 5(e) respectively. In the positive slope region, here also the carry propagation time of one full adder is smaller than the arrival inputs between each successive point in the positive slope region. Thus the RCA is again best suited in this positive slope region.

In the second region, we find four adders are working fast with only slight difference. They are CLA, CSLA, BCSA and BCSLA. The CLA is faster only up to middle of the second region (approximately 20 bits of 40 bits). The others perform faster with slight difference in the entire second region. Since the CLA is faster only half part of the second region, we can omit CLA in this region. So we have to choose one of the remaining adders. The BCSA performs faster than the CSLA and BCSLA in this region. But it leads to larger power and area. So we need to choose either CSLA or BCSLA. As explained earlier, the BCSLA is slightly slower than the CSLA, but it saves more area and power. So we suggest that the BCSLA is best in the entire region2. In the third region, the BCSA and BCLA perform faster with only slight differences. But the BCLA is faster than the BCSA in this region. Also the BCLA requires lesser area and power than the BCSA. So we can suggest BCLA be used for the third region.

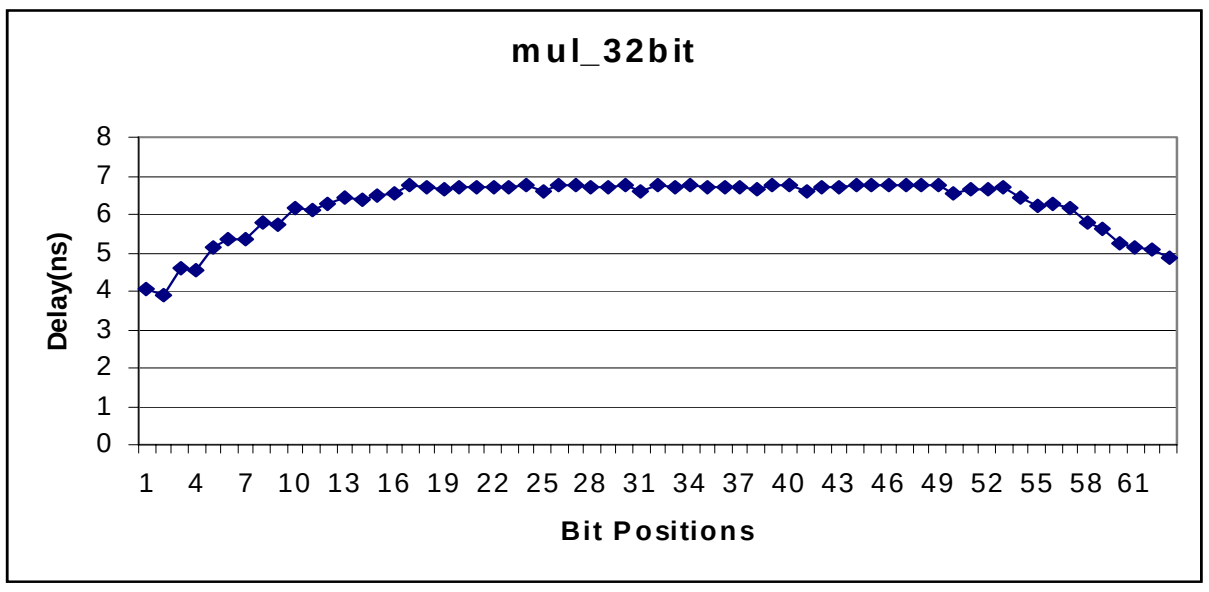

(a)

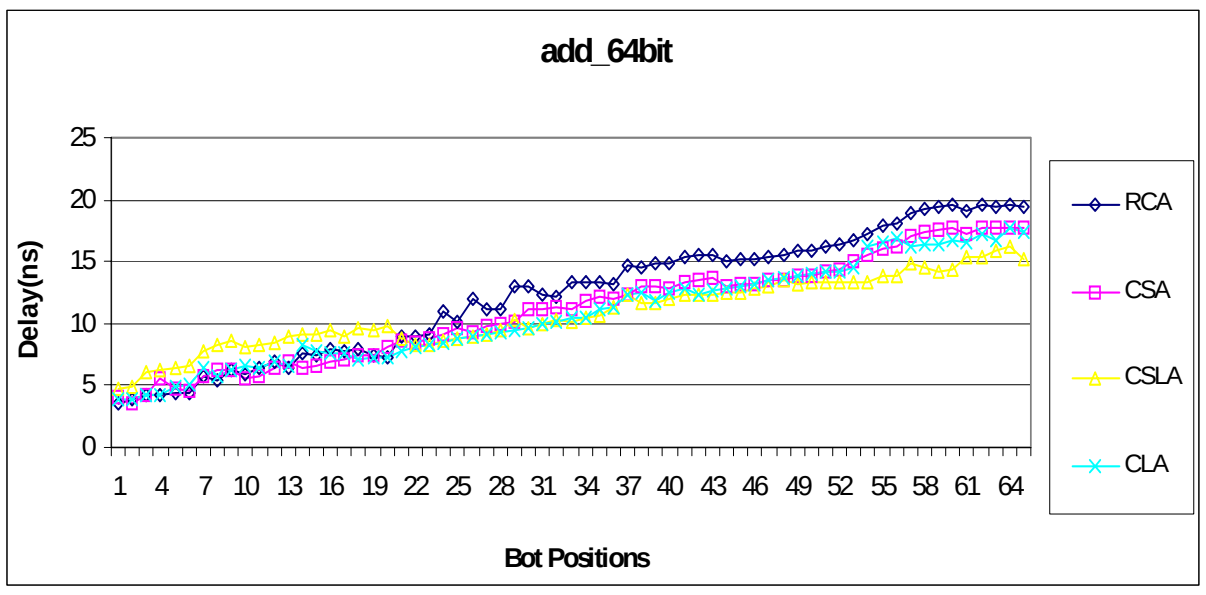

(b)

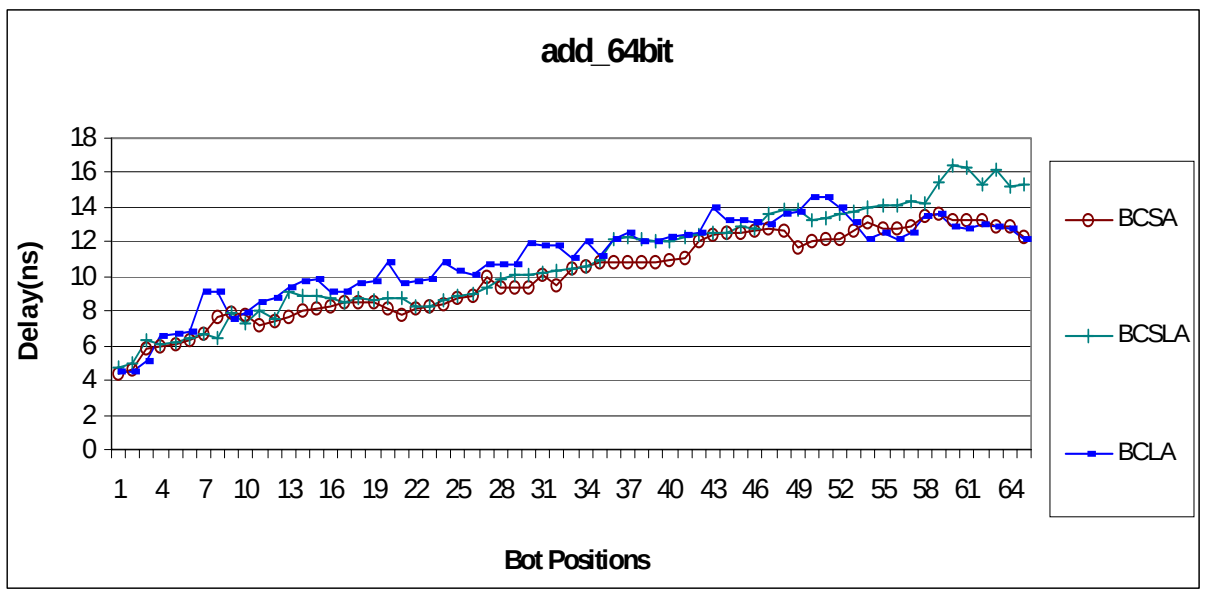

(c)

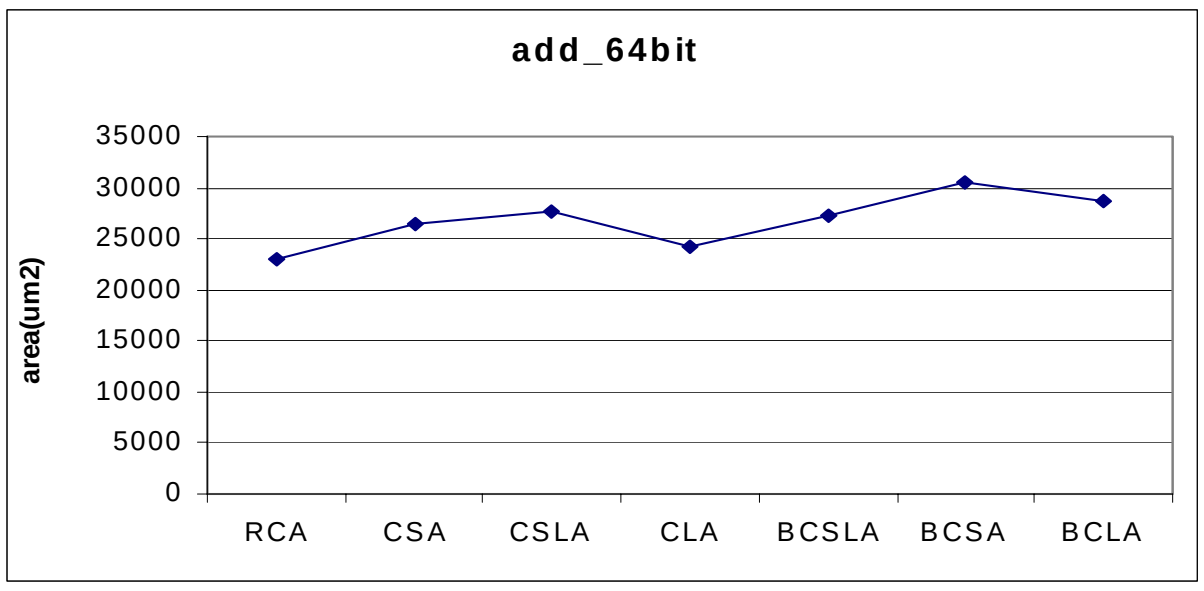

(d)

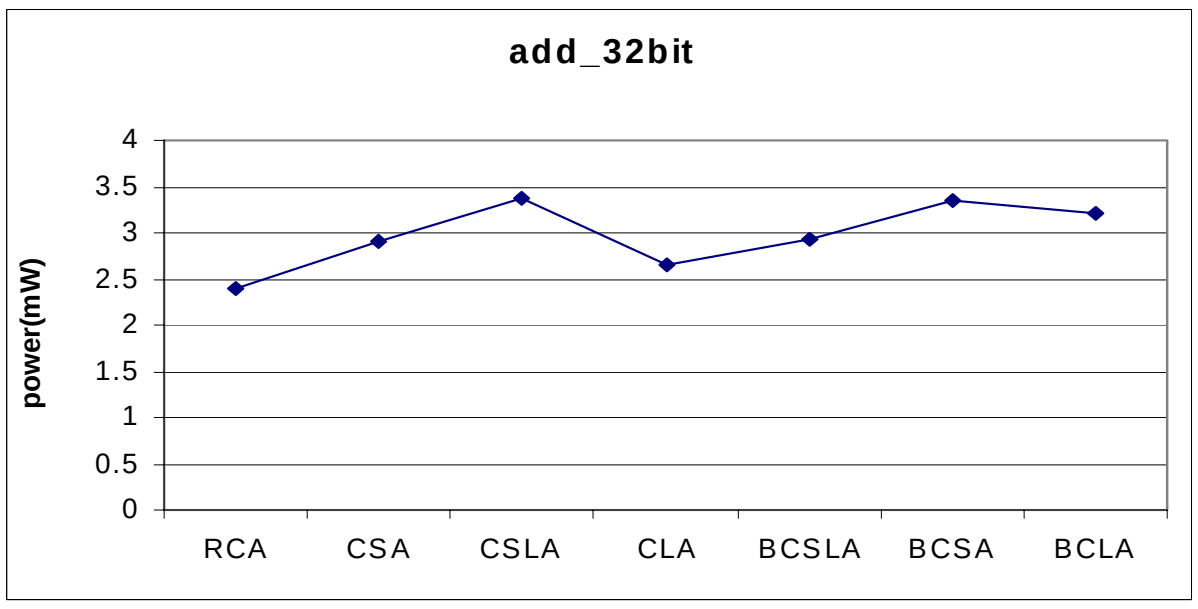

(e)

Fig. 5. Final adder analysis for 32 by 32 Dadda multiplier: (a). arrival profile of PPST, (b) delay analysis of RCA, CSA, CSLA and CLA, (c) delay analysis BCSA, BCSLA and BCLA, (d) area analysis, (e) power analysis.

### *D. 64 by 64 Dadda multiplier*

The input arrival profile to the final adder of 64 by 64 Dadda multiplier is shown in Fig. 6(a). Based on the arrival profile, here also we can divide this into 3 regions, in which the first region has a positive slope from the point 0 to 29 and the second region has a constant region from point 30 to 112 and the third region has a negative slope from the point 113 to 127.

The performance comparisons, of 7-types of 128-bit adders for the 64 by 64 multiplier, in terms of output timing are shown in Fig. 6(b) and Fig. 6(c). The area and power comparison is shown in Fig. 6(d) and Fig. 6(e) respectively. In the positive slope region, here also the carry propagation time of one full adder is smaller than the arrival inputs between each successive point in the positive slope region. Thus the RCA is again the best in this positive slope region. In the second region, we can find three adders are working fast with only slight difference. They are CSLA, BCSA and BCSLA. The CLA is faster only up to the quarter part of the second region (approximately 20 bits of 80 bits). The BCLA is faster from the last quarter part of the second region to the entire remaining third region. Since the effect of CLA and BCLA is on only certain parts of the second region, we have omitted these two adders in this region. The BCSA works faster than the CSLA and BCSLA in this region. However due to more area and power requirement of BCSLA, we have to choose one of the adders from CSLA and BCSLA.

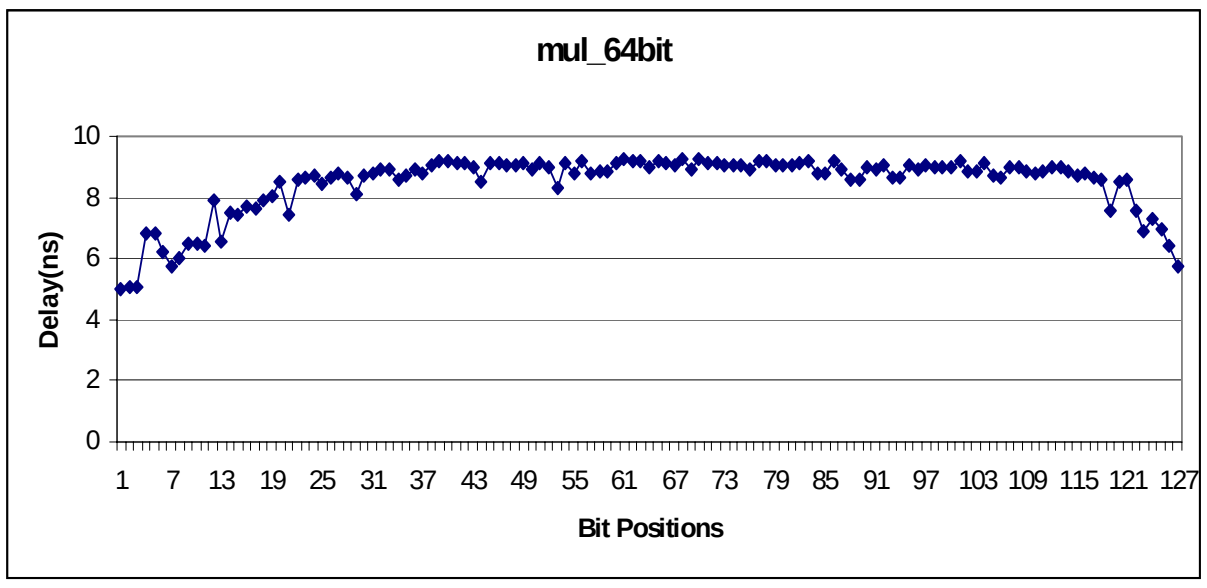

(a)

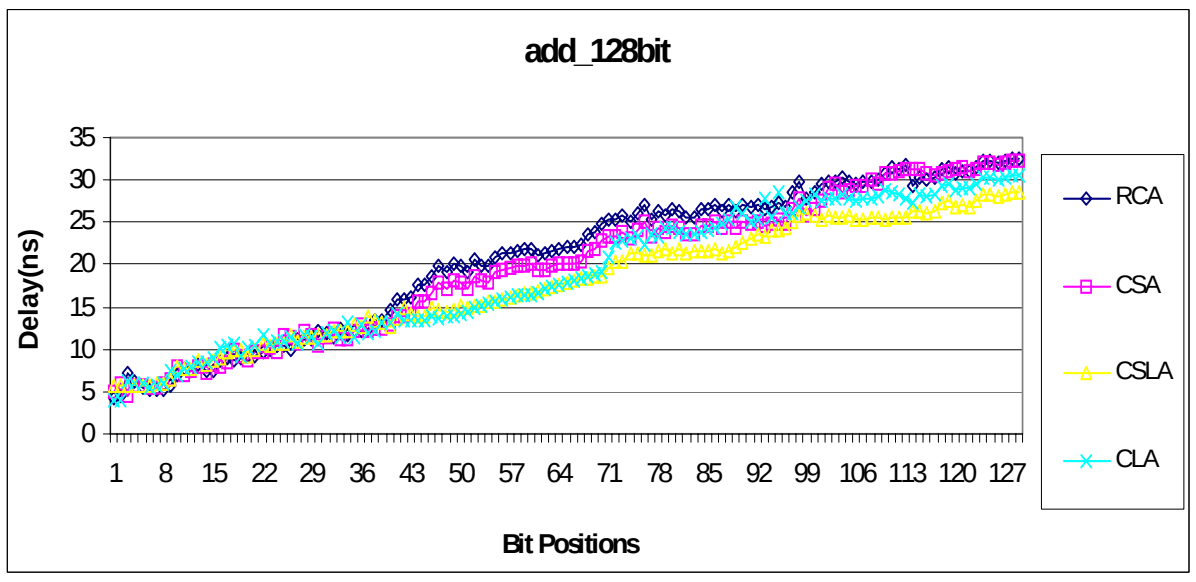

(b)

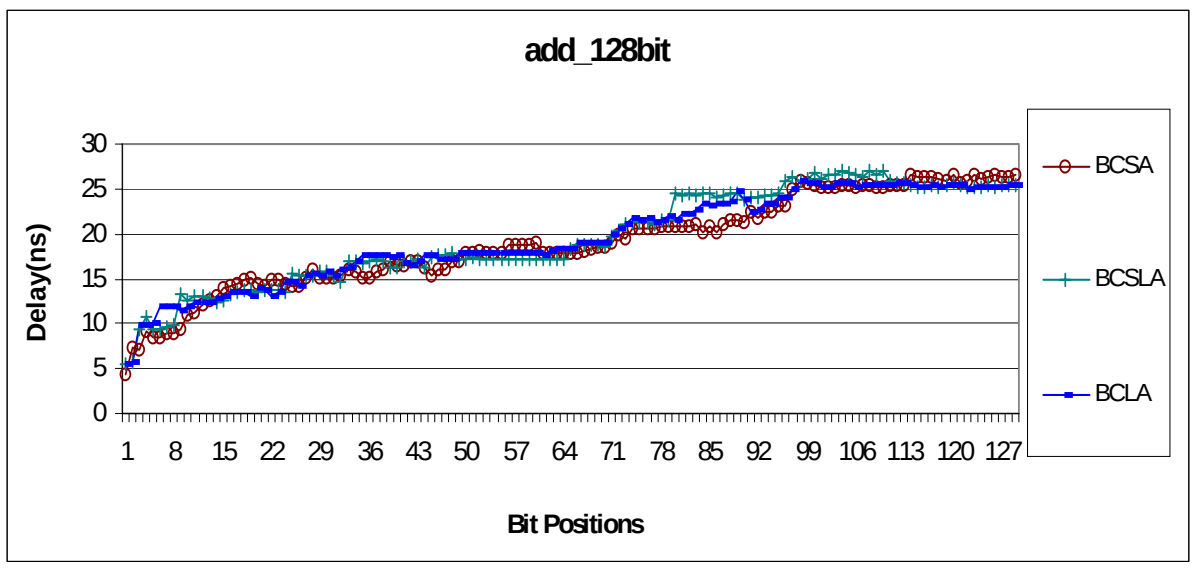

(c)

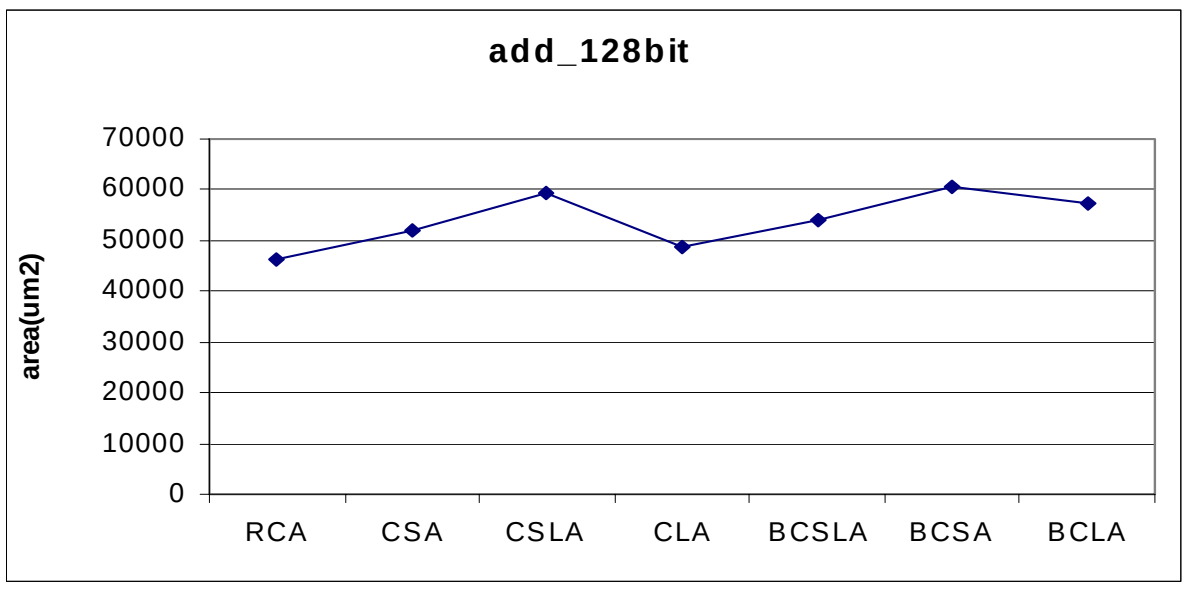

(d)

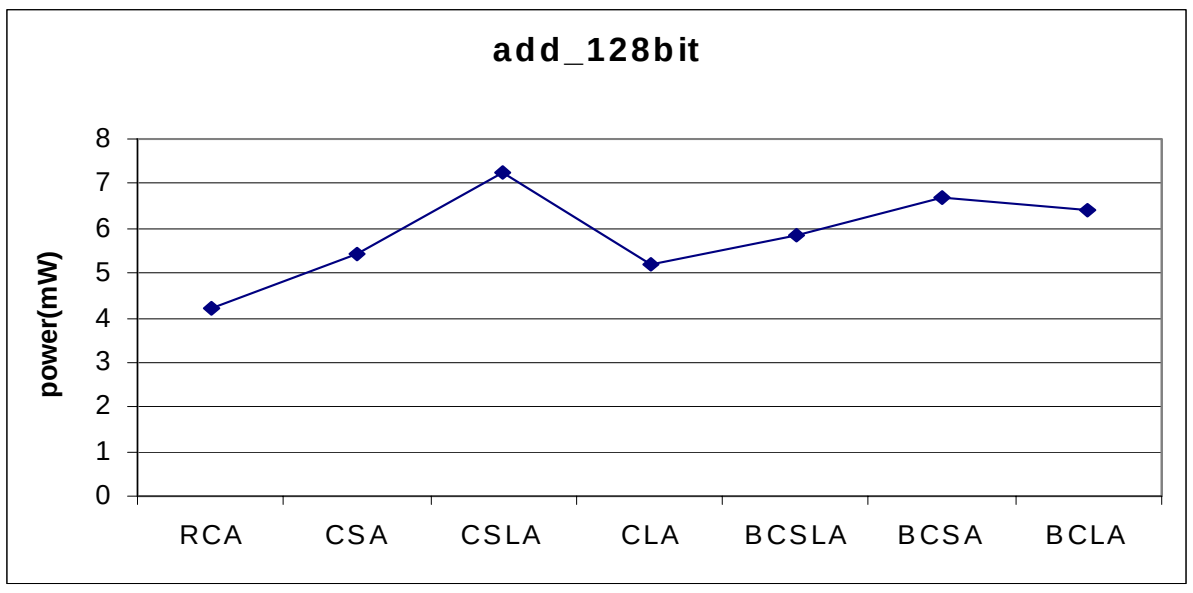

(e)

Fig. 6. Final adder analysis for 64 by 64 Dadda multiplier: (a). arrival profile of PPST, (b) delay analysis of RCA, CSA, CSLA and CLA, (c) delay analysis BCSA, BCSLA and BCLA, (d) area analysis, (e) power analysis.

As explained earlier, the BCSLA is slightly slower than the CSLA but it saves more area and power. So we can conclude that the BCSLA is suitable for this region.

In the third region, the BCSLA and BCLA perform faster with slight difference. But the BCLA is faster than the BCSLA in this region. Also the BCLA requires lesser area and power than the BCSA. So we choose BCLA for the third region.

## IV. EFFECTIVE PARTITIONING OF MULTIPLIER REGIONS

From the analysis of four operand size multipliers, first we can propose some simple equations to partition the 3 regions of the multiplier. The Table I (exact) is derived from the exact arrival bit positions to final adder. By shifting one or few bits from each region, we can develop equations for partitioning of the 3 regions. The Table II (approx) shows the contents of Table I (Exact) after modification of the bit width of each region.

From the Table I and Table II, we can conclude for *N*-bit multiplier (n must be >=4), the first region has ~n/2 bits, the second region has ~n+$2^x$ , where x = 0 for 4 to 7 (i.e., n to 2n-1), x = 1 for 8 to 15, etc. and the third region has ~n/4 bits. These expressions for the width of the three regions will reduce the design time in findings the three regions of the final adder.

TABLE I
NUMBER OF BITS IN 3 REGIONS (EXACT)

| Mul size | Region1 Bit Width | Region2 Bit width | Region3 Bit width |
|---|---|---|---|
| 8 | 0-5 | 6-14 | 15 |
| 16 | 0-6 | 7-25 | 26-31 |
| 32 | 0-15 | 16-52 | 53-63 |
| 64 | 0-29 | 30-112 | 113-127 |

TABLE II
NUMBER OF BITS IN 3 REGIONS (APPROX)

| Mul size | Region1 Bit width | Region2 Bit width | Region3 Bit width |
|---|---|---|---|
| 8 | 0-3=4 | 4-13 = 10 | 14-15=2 |
| 16 | 0-7=8 | 8-27 = 20 | 28-31=4 |
| 32 | 0-15=16 | 16-55 = 40 | 56-63=8 |
| 64 | 0-31=32 | 32-111 = 80 | 112-127=16 |

`

## V. FINAL ADDER DESIGN

From the analysis of 28 adders, the RCA is the best choice in the positive slope region. In the second region, there are three adders giving good performance. They are BCSA, BCSLA and BCLA. The overall adder structure will uses a hybrid adder structure, because each region uses a unique adder for giving better performance. If we make a hybrid structure in one region, the layout structure becomes more complex. So to avoid complexity, we can use a unique type of adder structure for each region. From the above analysis, the BCSLA gives optimal performance than the BCSA and BCLA. So the BCSLA is the best choice in the second region.

In the third region, both the BCSA and BCLA perform faster. Due to the large area and power requirement of BCSA, the BCLA is the best adder in this region. So we can choose the BCLA for the third region. Thus we arrive at the optimal final adder structure for parallel multiplier as shown in Fig. 7. The variable block BCSLA and BCLA is designed based on square-root method [15].

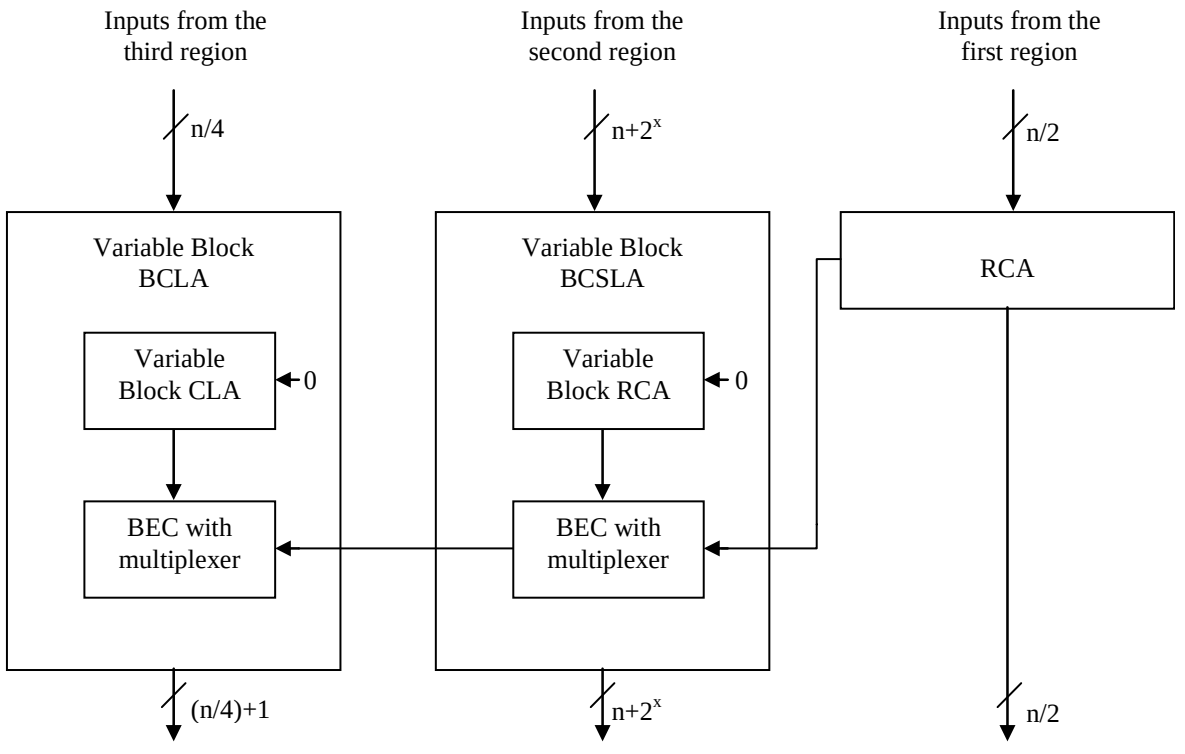

Fig. 7. Optimal final adder for all the three regions.

## VI. ASIC IMPLEMENTATION

The designs proposed in this paper have been developed using Verilog-HDL and synthesized in Cadence RTL compiler using typical libraries of TSMC 0.18um technology. The synthesized Verilog netlist and their respective design constraints file are imported to Cadence SoC Encounter and are used to generate automated layout from standard cells and placement & routing [16]. Parasitic extraction is performed using Encounter’s Native RC extraction tool. The extracted parasitic RC (SPEF format) is back annotated to Common Timing Engine in Encounter Platform and analyzed for static timing delay. The power analysis is done using Virtuso Ultrasim [17].

## VII. CONCLUSION

In this work we have obtained simple equations to obtain the bit size of the three different regions (positive slope, constant, negative slope) of the input arrival profile to the final CPA and analysis has been made on each region in terms of area, delay and power with various standard adders to suggest the structure of the optimal final adder for parallel multipliers. From the observed analysis results, the RCA, BCSLA & BCLA provide the optimal performance for positive slope region (width n/2), constant region (width 5n/4) and negative region (width n/4) respectively.

## ACKNOWLEDGMENT

This work was supported in part by the Integrated Circuit Design Laboratories, VIT University, Vellore, India.